\documentclass[
aps, 
prd, 
twocolumn, 
nofootinbib]{revtex4-1}

\usepackage{graphicx}%
\usepackage{dcolumn}%
\usepackage{bm}%
\usepackage{amsmath, amssymb, amsthm}
\usepackage{color}
\usepackage{adjustbox}
\usepackage[utf8]{inputenc}
\usepackage[english]{babel}
\usepackage{comment}
\usepackage{algorithm, algpseudocode}
\usepackage{booktabs}
\usepackage[caption=false]{subfig} %
\usepackage{float}
\usepackage{xcolor}
\usepackage{hyperref}
\hypersetup{
    colorlinks = true,
    linkcolor = {black},
    urlcolor = {blue},
    citecolor = {red},
}
\usepackage{url}
\usepackage{listings}
\usepackage{comment}
\usepackage{mathtools}
\usepackage{multirow}

\DeclareSymbolFont{bbold}{U}{bbold}{m}{n}
\DeclareSymbolFontAlphabet{\mathbbold}{bbold}

\newcommand{\papertitle}{The Karzas-Latter-Seiler Model of a High-Altitude Electromagnetic Pulse: \\
A New Numerical Code for an Old Model}

\begin{document}

\title{\papertitle}

\author{Gavin S. Hartnett\footnote{\texttt{\textup{hartnett@rand.org}}}}
\email{hartnett@rand.org}
\affiliation{
RAND Corporation \\
1776 Main St, Santa Monica, CA 90401
}

\begin{abstract}
A high-altitude nuclear blast can produce an electromagnetic pulse (EMP) capable of disrupting electronics on Earth. The basic phenomenology of the initial (E1) phase of the EMP was initially worked out in the 1960s by Longmire, Karzas, and Latter, and although more accurate and sophisticated EMP models have since been devised, the Karzas-Latter model is particularly simple and amenable to implementation as a numerical code. This paper accompanies the release of a new software implementation of an approximation of the Karzas-Latter model due to Seiler. This is, as far as we are aware, the only such publicly available numerical EMP code. After reviewing the physics and assumptions of the model, the numerical results for EMP simulations under a range of conditions are presented. It is shown how the results from multiple line of sight integrations of the field equations can be assembled to form a map of the EMP intensity across a broad geographic region. The model predictions are at least qualitatively correct and in general agreement with other simulation results, including the characteristic ``smile'' pattern in the spatial variation of the EMP intensity.
\end{abstract}

\maketitle

\section{Introduction \label{sec:intro}}
There is value to the public in developing open-source, transparent, and easily accessible models of the many Nuclear Weapons Effects (NWEs), such as the blast wave, radiation, fall-out, and the electromagnetic pulse (EMP). These models may be used to inform public discourse concerning the risk of nuclear war and can help in civil defense planning. For these purposes, it is not necessary to have particularly accurate models; even rough, order-of-magnitude estimates can be useful in educating the public and scenario planning. Fortunately, there are many simple and unclassified mathematical models of NWEs that could be developed into software tools.

\begin{figure}[!htbp]
    \centering
    \includegraphics[width=0.45\textwidth]{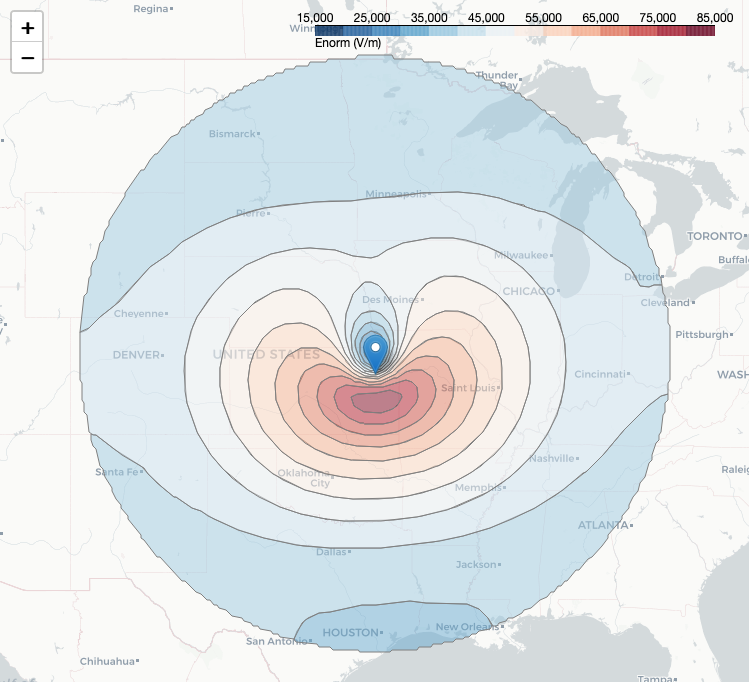}
    \caption{Contour plot of the maximum (over time) EMP intensity for a 5 kiloton nuclear detonation 100 km directly overhead Topeka, Kansas, USA.}
    \label{fig:Topeka_smile}
\end{figure}

The present work accompanies the public release of a Python implementation of the Karzas-Latter \cite{karzas1962electromagnetic, karzas1965detection} model for high-altitude electromagnetic pulse (HEMP).
\footnote{The code is available here: \href{https://github.com/gshartnett/karzas-latter-seiler}{https://github.com/gshartnett/karzas-latter-seiler}.}
This model treats the early E1 phase of the phenomenon and uses a series of rather crude approximations and simplifying assumptions. Nonetheless, we are not aware of any other public implementations of this model and believe that there is value in releasing it, provided its many limitations are well documented and communicated. In this work, we review the Karzas-Latter model of HEMP, as well as an extension of the model provided by Seiler \cite{seiler1975calculational} that greatly simplifies the calculation. We also discuss how this model, which was initially designed to compute the electric field strength along a line of sight from the burst to a target, may be extended to produce the characteristic ``smile diagrams" showing the magnitude of the EMP across a large region of the Earth's surface (Figure~\ref{fig:Topeka_smile}).

This paper is organized as follows. In Section~\ref{sec:karzas-latter} the Karzas-Latter model is briefly reviewed. Section~\ref{sec:results} presents the results from the numerical code, and Section~\ref{sec:discussion} concludes with a discussion of the limitations of this model. Additional details are relegated to two appendices; Appendix~\ref{app:seiler} reviews the Seiler approximation to the Karzas-Latter model, and Appendix~\ref{app:geomagnetic} contains details on the coordinate systems used as well as how the geomagnetic field has been modeled.

\subsection*{Related Work}
 
There is a long history of EMP modeling efforts. The first EMP model, CHAP, was developed by Longmire \cite{longley1972development}. Another model, HEMP, was developed in \cite{page1974electromagnetic}. CHAP serves as a standard comparison for newer models. For example, \cite{cui2003numerical, meng2013numerical} developed a FORTRAN code called MCHII, and the results seem to agree well with CHAP. Another fairly recent EMP model, EMPulse, was described in \cite{friedman2016empulse}. 

There have also been many efforts to go beyond the standard approach for treating the Maxwell equations and/or the source terms. Ref.~\cite{roussel2005prompt} demonstrated that the field equations are equivalent to those derivable from a Li\'{e}nard-Wiechert approach. Similarly, Ref.~\cite{eng2011development} compared the results from integrating Maxwell's equations (Jefimenko's equation) with CHAP. The multiple scattering of the primary Compton electrons was considered in \cite{farmer2015effect, farmer2016validity}. Scattered $\gamma$-rays were considered in \cite{li2020simulation}. An alternative approach to modeling HEMP, called the integral equation method, was developed in a series of papers \cite{zhang2017study, zhang2018using, zhang2020novel}. Finally, the HEMP caused by X-rays (rather than $\gamma$-rays) was studied in \cite{yao2021simulation}. Notably, none of the above works that developed numerical EMP codes appear to have made the code publicly available.

\section{The Karzas-Latter Model \label{sec:karzas-latter}}
Although all nuclear bursts generate an electromagnetic pulse (EMP), the EMP produced by the detonation of a nuclear weapon at high altitudes is greatly enhanced by the Earth's atmosphere and magnetic field \cite{glasstone1977effects}. There are three basic phases of high-altitude EMP - E1, E2, and E3. This work is solely focused on the E1 phase, whose basic phenomenology is as follows. First, the detonation releases a number of high-energy $\gamma$-rays which form a shell of radiation expanding outwards at the speed of light. (Prompt X-rays are also generated and contribute to the EMP through slightly different physics; this contribution will be ignored here). The downward traveling photons eventually enter Earth's atmosphere, where they interact with molecularly-bound electrons via Compton scattering to produce relativistic electrons that mostly travel coherently with the un-scattered photons. These so-called primary electrons then turn under the influence of Earth's magnetic field and begin to emit a pulse of synchrotron radiation
as they accelerate. As the electrons travel through Earth's atmosphere, they interact with air molecules and produce a large population of non-relativistic, positively-charged ions and so-called secondary electrons, with each primary Compton electron producing tens of thousands of secondary electrons. These charged particles form a conducting medium that generates a current that counteracts the pulse of cyclotron radiation. 

The physics of the E1 phase of EMP was mainly worked out by Longmire \cite{longmire1978electromagnetic} and Karzas-Latter (KL) \cite{karzas1962electromagnetic, karzas1965detection}. The KL model is the simplest model that captures the basic physics of the problem, and it is particularly amenable to numerical simulation. In this work, we will focus exclusively on the KL model and a later extension introduced by Seiler. A brief review of the model follows; more details may be found in the original articles, as well as \cite{seiler1975calculational, chapman1974computer}. 

The dynamics of the electromagnetic field are governed by the sourced Maxwell equations, and the KL model essentially amounts to a model of the source terms, together with a calculational approach to (approximately) solve the field equations. The sources consist of two components: a current $\bm{J}^C$ due to the motion of relativistic Compton (primary) electrons, and a current $\sigma \bm{E}$ due to the presence of non-relativistic secondary electrons. The contribution of molecular ions to the conductivity is neglected. The two currents are assumed to be non-zero only in the so-called absorption band of the atmosphere, which ranges from 20-50 km above the surface of the Earth. It is assumed that the Compton scattering which produces the primary electrons is entirely confined to this band. Outside of this band, the electromagnetic field propagates as in a vacuum, until it reaches a target point on the surface of the Earth.

The burst is assumed to occur at some point above the surface of the Earth. Spherical coordinates centered around the burst point are used, with the $z$-axis aligned with the local orientation of the geomagnetic field. It is convenient to work in terms of the retarded time coordinate $\tau \equiv t - r/c$. Thus, the coordinates used are: $(\tau, r, \theta, \phi)$, and these should not be confused with a spherical coordinate system centered on the Earth. The field equations to be solved are:
\begin{subequations}
\label{eq:maxwell}
\begin{equation}
    \frac{2}{c} \frac{1}{r} \partial_r (r E_{\theta}) + \mu_0 J_{\theta}^C + \mu_0 \sigma E_{\theta} = 0 \,,
\end{equation}
\begin{equation}
    \frac{2}{c} \frac{1}{r} \partial_r (r E_{\phi}) + \mu_0 J_{\phi}^C + \mu_0 \sigma E_{\phi} = 0 \,.
\end{equation}
\end{subequations}
These equations are derived by starting from sourced electromagnetic wave equation for $\bm{E}$, converting to retarded coordinates, and then making the Karzas-Latter approximation wherein the spatial variation is assumed slow compared to the time variation. The radial component has been neglected, as it is non-zero only in the absorption band. A similar set of equations exist for the radiative magnetic field, and in fact, it can be shown that $B_{\theta} = -E_{\phi}$ and $B_{\phi} = E_{\theta}$ \cite{karzas1965detection}. Outside the absorption band, the source terms are zero, and $E_{\theta}, E_{\phi} \propto 1/r$, with the constant of proportionality given by the value at the edge of the band. The total EMP intensity at a radius $r > r_{\text{max}}$ is therefore given by:
\begin{equation}
	{E(r) = \sqrt{E_{\theta}^2(r_{\text{max}}) + E_{\phi}^2(r_{\text{max}})} \frac{ r_{\text{max}} }{ r } } \,.\end{equation}

To fully specify the model, the source terms are needed. The components of the Compton (primary) current are:
\begin{widetext}
\begin{subequations}
\label{eq:current}
    \begin{equation}
    J_{\theta}^C(\tau, \bm{r}) = 
    \begin{dcases}
    - e g(\bm{r}) V_0 \sin\theta \cos\theta \int_0^{R(\bm{r})/V_0} d\tau' f\left( \tilde{\tau}(\tau, \tau') \right) (\cos(\omega \tau') - 1)  \,, & r_{\text{min}} < r < r_{\text{max}} \,, \\
    0 \,, & \text{else} \,.
    \end{dcases}
\end{equation}
\begin{equation}
    J_{\phi}^C(\tau, \bm{r}) =
    \begin{dcases}
    - e g(\bm{r}) V_0 \sin\theta \int_0^{R(\bm{r})/V_0} d\tau' f\left( \tilde{\tau}(\tau, \tau') \right) \sin(\omega \tau') \,, & r_{\text{min}} < r < r_{\text{max}} \,, \\
    0 \,, & \text{else} \,.
    \end{dcases}
\end{equation}
\end{subequations}
and the conductivity of the secondary electrons is
\begin{equation}
\label{eq:conductivity}
    \sigma(\tau, \bm{r}, \bm{E}) =
    \begin{dcases}
    \frac{e^2 q V_0 g(\bm{r})}{m R(\bm{r}) \nu_c(\tau, \bm{E})} \int_{-\infty}^{\tau} d \tau' \int_0^{R(\bm{r})/V_0} d\tau'' f\left( \tilde{\tau}(\tau', \tau'') \right) \,, & r_{\text{min}} < r < r_{\text{max}} \,, \\
    0 \,, & \text{else} \,. 
    \end{dcases}
\end{equation}
\end{widetext}
The quantities that appear in these expressions are as follows. $r_{\text{max}}, r_{\text{min}}$ are the radii of the upper and lower boundaries of the absorption band. The electron charge and mass are $e$, $m_e$, respectively. The velocity of the Compton electrons is $V_0$, $\beta = V_0/c$ is the velocity ratio, and $\gamma = 1/\sqrt{1 - \beta^2}$ is the Lorentz factor. The average number of secondary electrons generated by a single primary electron is $q$. The cyclotron frequency of the Compton electrons is $\omega = e B_E/\gamma m_e$, and $B_E$ is the magnitude of the Earth's magnetic field (assumed constant along the line of sight). The normalized pulse function $f$ quantifies the fractional amount of radiation output of the burst as a function of time. The argument appearing in the pulse function is
\begin{equation}
    \tilde{\tau}(\tau, \tau') = \tau - (1- \beta \cos^2\theta)\tau' + \beta \sin^2\theta \frac{\sin(\omega \tau')}{\omega} \,.
\end{equation}
Recall that the $z$-axis has been aligned with the geomagnetic field, so that $\theta$ is both the polar angle and the angle between the radial direction of motion and the geomagnetic field.

The rate function $g$, which determines the local rate at which Compton (primary) electrons are produced at different radii, is 
\begin{equation}
    g(\bm{r}) = \frac{Y_{\gamma}}{K} \frac{\exp\left(- \int_0^r \frac{\mathrm{d} r' }{\lambda(\bm{r}')} \right)}{ 4 \pi r^2 \lambda(\bm{r}) } \,.
\end{equation}
Here $Y_{\gamma}$ is the $\gamma$-ray yield, measured in units of energy, $K$ is the kinetic energy of the Compton (primary) electrons, $\lambda$ is the mean free path for $\gamma$-rays interacting with the Earth's atmosphere via Compton scattering.

\begin{figure*}
    \centering
    \includegraphics[width=0.95\textwidth]{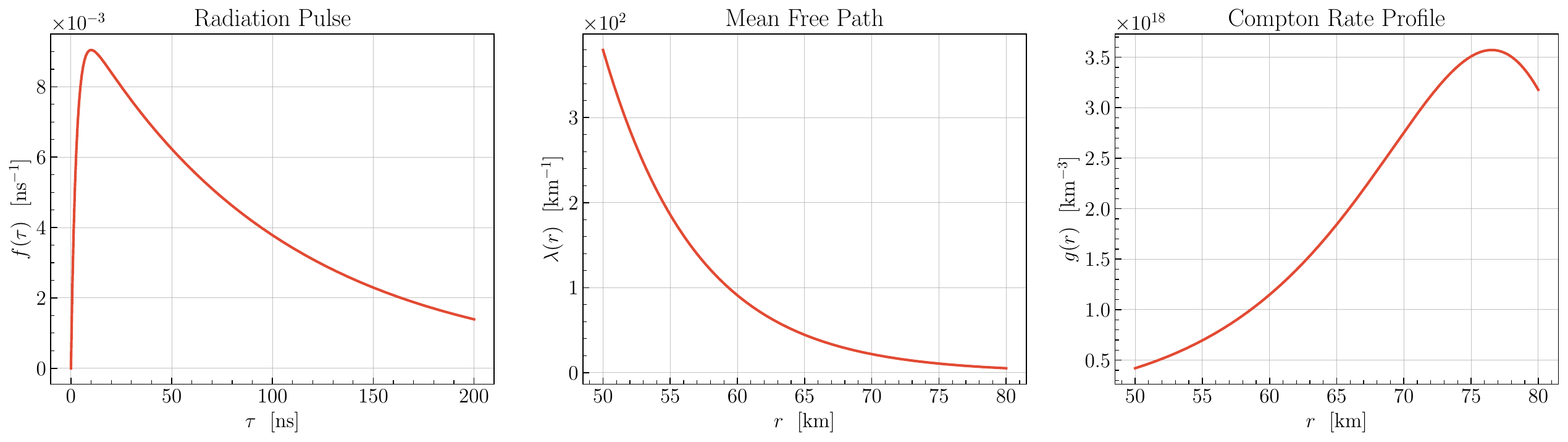}
    \caption{(\textit{Left}): The radiation pulse profile, Eq.~\ref{eq:pulse}, with parameters $a = 0.01$ ns and $b = 0.37$ ns. \textit{(Center)}: The mean free path as a function of distance from the burst. (\textit{Right}): The Compton rate profile as a function of distance from the burst. All plots were made following Seiler, as detailed in Appendix~\ref{app:seiler} and using the default parameter values listed in Table~\ref{table:parameters}.}
    \label{fig:MFP_and_ComptonRate}
\end{figure*}

The integral in the argument of the exponential term is over the radial coordinate and extends from the burst point $r=0$ to an arbitrary radius $r$ away from the blast. Finally, $R(\bm{r})$ is the range of the primary (Compton) electrons (it is assumed that the primary electrons travel at a constant velocity $V_0$ until abruptly coming to a stop after traveling a distance $R(\bm{r})$), and $\nu_c(\tau, \bm{E})$ is the electron collision frequency. Following \cite{karzas1965detection} we have made the additional assumption that the electron collision frequency is large relative to the frequencies in the conductivity and electric field. Typical curves for the radiation pulse, mean free path, and Compton rate profile are plotted in Figure~\ref{fig:MFP_and_ComptonRate}.

Naively, the problem of modeling the EMP entails solving a (3+1)-dimensional system of PDEs. The assumptions and approximations made by Karzas-Latter have greatly reduced the complexity, simplifying the problem to a system of two coupled ODEs, Eq.~\ref{eq:maxwell}. The time-dependent source terms, Eq.~\ref{eq:current},~\ref{eq:conductivity}, are given in terms of integrals which may be computed numerically. Thus, the solution of the field equations corresponds to the electric field configuration at a single moment in retarded time $\tau$, along a line of sight from the burst point to a target point.

To complete the specification of the model, additional expressions for quantities such as the pulse function or air density must be given. Unless stated otherwise, we will follow Seiler \cite{seiler1975calculational}. The pulse function is modeled as the difference of exponentials:
\begin{equation}
    \label{eq:pulse}
    f(t) = \begin{cases}
    		0 \,, & t < 0  \\
    		\frac{a b}{b - a} \left( e^{- a t} - e^{-b t} \right) \,, & t \ge 0 \,.
    		\end{cases}
\end{equation}
Here $a, b$ are shape parameters. The function is normalized such that $\int_{-\infty}^{\infty} \mathrm{d} t \, f(t) = 1$. 

The air density is modeled as an exponential atmosphere:

\begin{equation}
    \rho(\bm{r}) = \rho_0 \exp\left(-z/S\right) \,,
\end{equation}

where $z$ is the altitude measured with $z=0$ corresponding to the Earth's surface, and $S$ is the scale height. The altitude may be given in terms of the radius from the burst point as: $z = H - r \cos(A)$, where $H$ is the burst height and $A$ is the angle between the radial vector and the normal vector to the Earth's surface.
For the scale height, the code will use a default value of $S = 7$ km. 

Many quantities are inversely proportional to the air density, such as the mean free path for $\gamma$-rays interacting with the Earth's atmosphere via Compton scattering:

\begin{equation}
    \lambda(\bm{r}) = \lambda_0 \frac{\rho_0}{\rho(\bm{r})} \,.
\end{equation}

Here a value of $\lambda_0 = 0.3$ km is used. The Compton range also scales inversely with atmospheric density:

\begin{equation}
	\label{eq:range}
	R(\bm{r}) = R_0 \frac{\rho_0}{\rho(\bm{r})} \,, \qquad R_0 = \frac{4.12}{\rho_0} K^{\left(1.265 - 0.9954 \ln K \right)} \,.
\end{equation}

Here the units are implicit: $K$ is measured in MeV, $\rho_0$ in kg/m$^3$, and $R_0$ in m. The lifetime of the Compton (primary) electrons is given by $R(\bm{r})/V_0$, and this quantity appears as the upper limit of integration in the source term integrals Eq.~\ref{eq:current}~\ref{eq:conductivity}. Seiler uses a maximum value of 1 microsecond for the lifetime, which roughly corresponds to setting $R(\bm{r}) \rightarrow \min\left( R(\bm{r}), 300 \text{ m} \right)$.

The secondary electron collision frequency also scales inversely with the atmospheric density: ${\nu_c(\tau, \bm{E}) = \nu_{c0}(\tau, \bm{E}) \rho_0/\rho(\bm{r})}$, with the sea-level value given by:        
\begin{equation}
	\label{eq:collisionfreq}
	\nu_{c0}(\tau, \bm{E}) = \min(4400, \max(\nu_{c0}^a, \nu_{c0}^b, \nu_{c0}^c) ) \,, 
\end{equation}
where
\begin{subequations}
\begin{equation}
	\nu_{c0}^a = -250 \, \tau + 4500 \,,
\end{equation}
\begin{equation}
	\nu_{c0}^b = 
	\begin{cases}
	0.043 \, E + 1600 & \text{if } E < 50,000 \,, \\
	0.06 \, E + 8000 & \text{otherwise} \,,
	\end{cases}
\end{equation}
\begin{equation}
	\nu_{c0}^c = 2800 \,.
\end{equation}
\end{subequations}
Here $\tau$ is measured in nanoseconds, $E$ in V/m, and $\nu_C$ has units of ns$^{-1}$. Note that $\nu_c(\tau, \bm{E})$ implicitly depends on the position $\bm{r}$ both through the air density factor and the $E$-field.

Lastly, the kinetic energy of the primary electrons is needed. For Compton scattering, the final $E_{\gamma, f}$ and initial $E_{\gamma, i}$ $\gamma$-ray energies are related via
\begin{equation}
    E_{\gamma,f} = \frac{ E_{\gamma, i}}{1 + (1-\cos\theta_C) E_{\gamma, i}/(m_e c^2)} \,.
\end{equation}
The electron kinetic energy is then given by ${K = E_{\gamma,f} - E_{\gamma,i}}$, which determines the velocity via: ${\beta = \sqrt{(K^2 + 2 K m_e c^2)/(K + m_e c^2)^2}}$. A common choice in the literature is to assume that the $\gamma$-rays have an energy of 1.5 MeV, and that the electrons are created with the maximum energy possible through Compton scattering (corresponding to a scattering angle of $\theta_C = \pi$). This results in an electron kinetic energy of 1.28 MeV. Finally, it is assumed that each secondary electron has a kinetic energy of 33 eV, and so the average number of secondary electrons per primary is $q = K/33$ eV.

The full list of parameters (not including constants of nature) is summarized in Table~\ref{table:parameters}. The parameters are divided into two categories: ``fundamental'' parameters such as the height of burst $H$ or the angle $A$, and ``derivative'' parameters such as the distance from the burst to a ground target: $r_{\text{target}} = H/\cos(A)$. The fundamental/derivative terminology is used to convey the fact that the second set of parameter values are derived from the first ones. Unless stated otherwise, the default values are used for all results discussed below in Section~\ref{sec:results}.

\begin{table*}[htbp]
\captionsetup{justification=centering}
\caption{\label{table:parameters} Fundamental and Derivative Model Parameters}
\begin{tabular}{lll}
	\hline
    \multicolumn{1}{l}{Parameter} & \multicolumn{1}{l}{Meaning} & \multicolumn{1}{l}{Default Value} \\ \hline
    \multicolumn{1}{l}{$a$} & pulse function parameter & $0.01$ ns$^{-1}$ \\
    \multicolumn{1}{l}{$b$} & pulse function parameter & $0.37$ ns$^{-1}$ \\
    \multicolumn{1}{l}{$K$} & primary electron kinetic energy & $1.28$ MeV \\
    \multicolumn{1}{l}{$S$} & atmosphere scale height & $7$ km \\
    \multicolumn{1}{l}{$\rho_0$} & air density at sea level & $1.293$ kg m$^{-3}$ \\
    \multicolumn{1}{l}{$B_E$} & magnitude of geomagnetic field & $3 \times 10^{-5}$ T \\
    \multicolumn{1}{l}{$\lambda_0$} & $\gamma$-ray mean free path at sea level & $0.3$ km \\
    \multicolumn{1}{l}{$H$} & height of burst & $100$ km \\
    \multicolumn{1}{l}{$A$} & angle between line of sight and Earth normal & $0$ radians \\
    \multicolumn{1}{l}{$Y_{\gamma}$} & $\gamma$ radiation blast yield & $0.25$ kt \\
	\\ \hline
    \multicolumn{1}{l}{Parameter} & \multicolumn{1}{l}{Meaning} & \multicolumn{1}{l}{Default Value} \\ \hline
    \multicolumn{1}{l}{$R_0$} & primary electron range & (see Eq.~\ref{eq:range})  m\\
    \multicolumn{1}{l}{$V_0$} & primary electron velocity & $c \sqrt{(K^2 + 2 K m_e c^2)/(K + m_e c^2)^2}$ m s$^{-1}$ \\
    \multicolumn{1}{l}{$\beta$} & normalized primary electron velocity & $V_0/c$ \\
    \multicolumn{1}{l}{$\gamma$} & primary Lorentz factor & $(1 - \beta^2)^{-1/2}$ \\
    \multicolumn{1}{l}{$q$} & average number of secondary electrons & $K/(33$ eV) \\    
    \multicolumn{1}{l}{$\omega$} & primary electron cyclotron frequency & $e B_E/\gamma m_e$ s$^{-1}$ \\
    \multicolumn{1}{l}{$\theta$} & angle b/w radial direction and geomagnetic field  & $\pi/2$ radians \\
    \multicolumn{1}{l}{$r_{\text{min}}$} & radius of upper absorption layer boundary & $(H - 50\text{ km})/\cos(A)$ km \\
    \multicolumn{1}{l}{$r_{\text{max}}$} & radius of lower absorption layer boundary & $(H - 20\text{ km})/\cos(A)$ km \\
    \multicolumn{1}{l}{$r_{\text{target}}$} & radius to ground-based target & $H/\cos(A)$ km \\
    \\ \hline 
\end{tabular}
\end{table*}

The Karzas-Latter model makes several approximations and assumptions that should be noted. The simplified form of the field equations (Eq.~\ref{eq:maxwell}) is due to the high-frequency approximation wherein the spatial variation is assumed to be slow compared to the time variation. Additionally, the positive ions created along with the secondary electrons are not modeled at all, and both species of electrons are treated very crudely. For example, in the KL model, each primary electron has the same energy, is moving in the same direction, and produces the same number of secondary electrons. Also, the primary electrons should slow down continuously as they ionize the atmosphere; instead, they are assumed to travel at a constant velocity until coming to an abrupt stop after an altitude-dependent stopping distance is traversed. 

Another key limitation of the model is that it ignores backreaction -- for example, the turning of the primary electrons in Earth's magnetic field will generate an additional magnetic field contribution, which will then affect the turning, which will then affect the generated field, and so on. Incorporating this backreaction is also called self-consistency in the literature. Although the KL model is not self-consistent with respect to the magnetic field, it at least incorporates some aspect of the backreaction due to the $\theta$- and $\phi$-components of the electric field in that conductivity of the secondary electrons does take into account the current value of the electric field via the electron collision frequency, Eq.~\ref{eq:collisionfreq}. However, note that the contribution of the radial component is neglected here. (The radial component is zero outside the absorption band, but non-zero in the absorption band, and it would therefore contribute to the total field strength.) Finally, the model also assumes that the geomagnetic field is constant along the line of sight. 

Even with these many assumptions, the model still involves solving integro-ODEs. The conductivity, Eq.~\ref{eq:conductivity} is especially challenging as it contains two nested integrals. These considerations motivated Seiler to introduce a further approximation to the model based on a small-time expansion of the source terms \cite{seiler1975calculational}, which is reviewed in Appendix~\ref{app:seiler}. We have implemented the resulting model, which we refer to as the Karzas-Latter-Seiler model, in a Python code, and in the next section we present some numerical results obtained using the code.

\section{Numerical Results \label{sec:results}}
\begin{figure}[!htbp]
    \centering
    \includegraphics[width=0.45\textwidth]{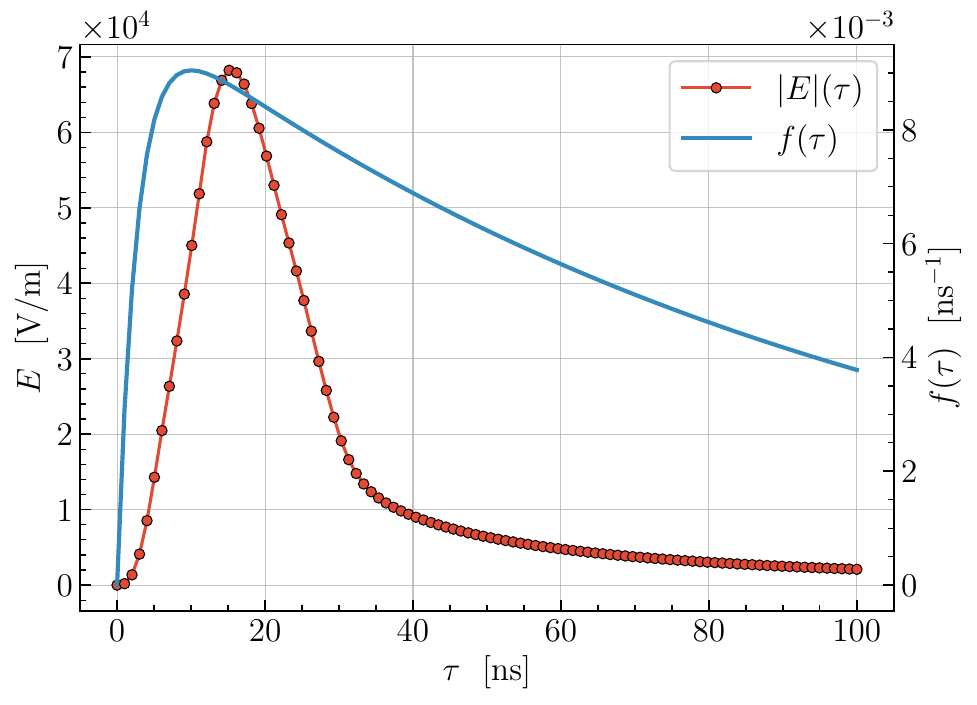}
    \caption{Magnitude of the electric field at the Earth's surface (red curve with data points, left-axis) and the $\gamma$-ray pulse (blue curve, solid line, right-axis), both as functions of retarded time. Note that the peak electric field occurs after the peak pulse value.}
	\label{fig:EMP_and_gamma_pulse}
\end{figure}

The output of the model is the electric field along the line of sight from the burst to the target. Of primary interest is the value of the electric field evaluated at the target point, typically located on the surface of the Earth. The temporal variation is easily calculated by solving the model for different values of retarded time. An example is shown in Figure~\ref{fig:EMP_and_gamma_pulse} for a burst with the default parameter values, as listed in Table~\ref{table:parameters}. The field quickly rises to a peak value in ~15 ns, and then tapers off with a long tail. 

\begin{figure}[ht]
    \centering
    \includegraphics[width=0.5\textwidth]{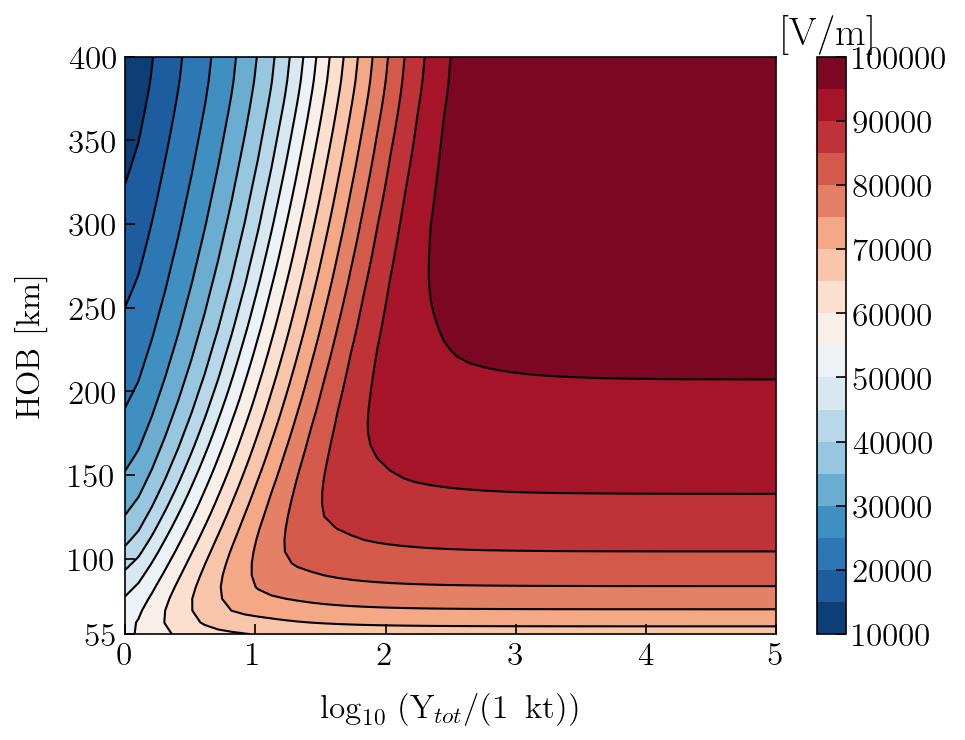}
    \caption{
Variation of the maximum EMP intensity at ground zero, $|\bm{E}|$, as both the height of burst (HOB) and the total yield $Y_{\text{tot}}$ is varied.
}
    \label{fig:HOB_yield_scan.png}
\end{figure}

Next, in Figure~\ref{fig:HOB_yield_scan.png} the variation of the EMP intensity as both the HOB and the weapon yield are varied is shown, with the yield ranging across 5 orders of magnitude from 1 kt to 100 Mt. For reference, the yield of the weapon dropped on Hiroshima has been estimated to be 15 kt, and the yield of the largest weapon ever detonated, the Tsar Bomba, has been estimated to be roughly 60 Mt. The plot shows that increasing the yield will increase the intensity, but with diminishing gains. In particular, any further increases beyond a few hundred kt have a very small effect on the peak EMP intensity. An important practical consideration that arises for large yields is that the Maxwell field equations, Eq.~\ref{eq:maxwell}, become stiff, which motivates the use of an implicit numerical integration scheme.

It is often useful in policy discussions to understand how the intensity of the EMP varies over the surface of the Earth. Although the Karzas-Latter-Seiler model is a 1-dimensional code, the results of many different line of sight integrations can be stitched together to form a full solution to the (3+1)-dimensional problem. First, as above the variation over time can be easily incorporated by solving the model for different values of retarded time. To incorporate the additional two spatial dimensions, many different line of sight integrations can be considered, with each line of sight vector extending from the same burst point to a different target point. For each target point the model will receive as input a spatially-dependent value of $A$ (the angle between the line of sight and the vertical), $B_E$ (the norm of the magnetic field), and $\theta$ (the angle between the line of sight and the magnetic field), and will output a spatially-dependent profile for the EMP intensity $E(\tau) = |\bm{E}(\tau)|$. Given the many assumptions made thus far, it is reasonable to model the geomagnetic field as a simple dipole. However, the code also supports the far more accurate International Geomagnetic Reference Field (IGRF) model of the geomagnetic field \cite{alken2021international}). Additional details are provided in Appendix~\ref{app:geomagnetic}.

Figure~\ref{fig:Topeka_smile} shows a contour plot for the maximum (in time) EMP intensity as a function of latitude and longitude for a 5 kiloton burst point 100 km directly overhead Topeka, Kansas, USA. The characteristic ``smile'' pattern is clearly present -- the intensity is comparatively weak at ground zero, and weaker still directly North of ground zero, but is quite large in a half-annular region immediately South of ground zero. The EMP only affects ground points that are within a line of sight from the burst, which limits the effect to a circular region whose extent is determined by the burst height. Thus, the surface intensity exhibits a discontinuous drop-off to zero outside this boundary.

\begin{figure*}[!htbp]
    \centering
    \includegraphics[width=0.95\textwidth]{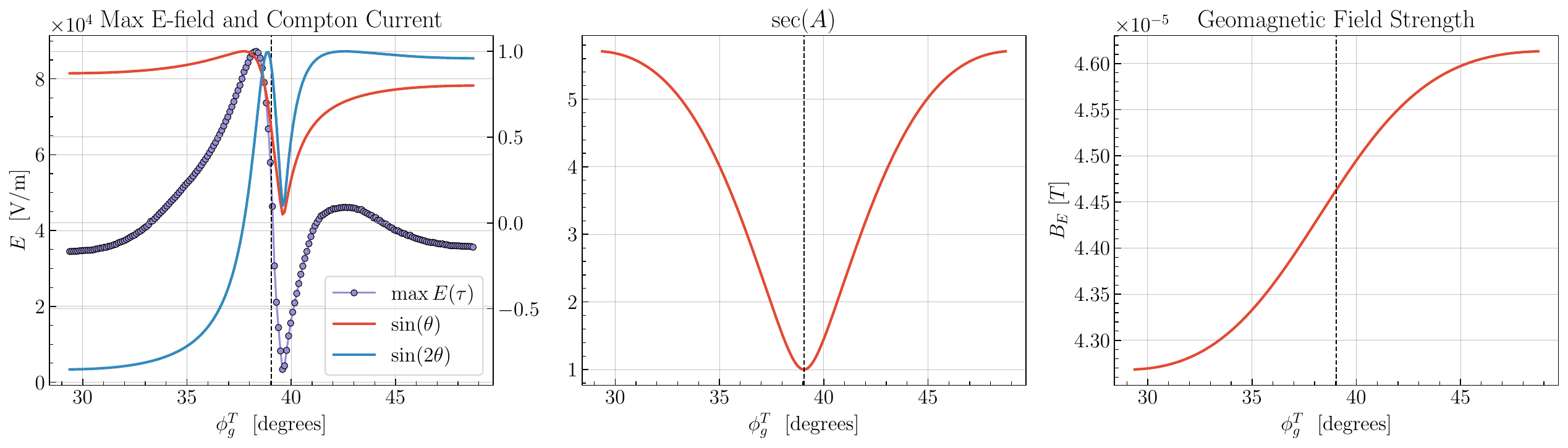}
    \caption{(\textit{Left}): The variation of the max (in time) $E$-field intensity (left axis), and $\sin(\theta)$ and $\sin(2\theta)$ (right axis) with target latitude. Note that the max $E$-field intensity occurs between the max intensities of either current component. (\textit{Center}): The variation of $\sec(A)$ with target latitude (the distance to a ground target varies as $\sec(A)$). (\textit{Right}): The variation of $B_E$ with target latitude. Ground zero is indicated by a dashed vertical line. All plots correspond to a burst point located 100 km directly overhead Topeka, Kansas.}
	\label{fig:angle_variation_with_latitude}
\end{figure*}

The observed variation of EMP intensity with latitude can be understood in terms of a number of geometrical effects. Of foremost importance is the angle between the line of sight and the geomagnetic field, $\theta$. The current components vary with $\theta$ as $J_{\theta}^C \propto \sin(2\theta)$ and $J_{\phi}^{C} \propto \sin(\theta)$, c.f. Eq.~\ref{eq:current}, and so a line of sight with $\theta=0$ will be entirely protected from the EMP as the currents (and therefore $E$) will be precisely zero. For a burst in the Northern hemisphere, $\theta$ exhibits a sharp decrease to zero immediately North of ground zero (GZ), and a sharp increase immediately South, implying that the EMP will attain a minimal value of zero just North of GZ and a maximal value just South of GZ. Note that the two current components are maximized at different angles, $J_{\theta}^C$ at $\theta = \pi/4$ and $J_{\phi}^C$ at $\theta = \pi/2$, and therefore the angle at which the EMP intensity is maximized can be expected to fall between these two values. A second geometrical effect is that distance to the ground target is $r = H / \cos(A)$, and thus the $1/r$ fall-off experienced by the $E$-field varies as $\sec(A)$. This increases with distance away from GZ, and the net effect is to both dampen the peak intensity of the smile and to move the curve of the smile to closer to GZ. These effects are discussed at greater length in \cite{savage2010early}. A third effect comes from the angular dependence of the norm of the geomagnetic field. The norm of the dipole magnetic field increases by a factor of 2 as one moves from the (magnetic) equator to the (magnetic) poles. The variations of the max (in time) $E$-field intensity, $\sin(\theta)$, $\sin(2\theta)$, $\sec(A)$, and $B_E$ are shown in Figure~\ref{fig:angle_variation_with_latitude} for a burst point directly overhead Topeka, Kansas. 

\begin{figure*}[!htbp]
    \centering
    \includegraphics[width=0.95\textwidth]{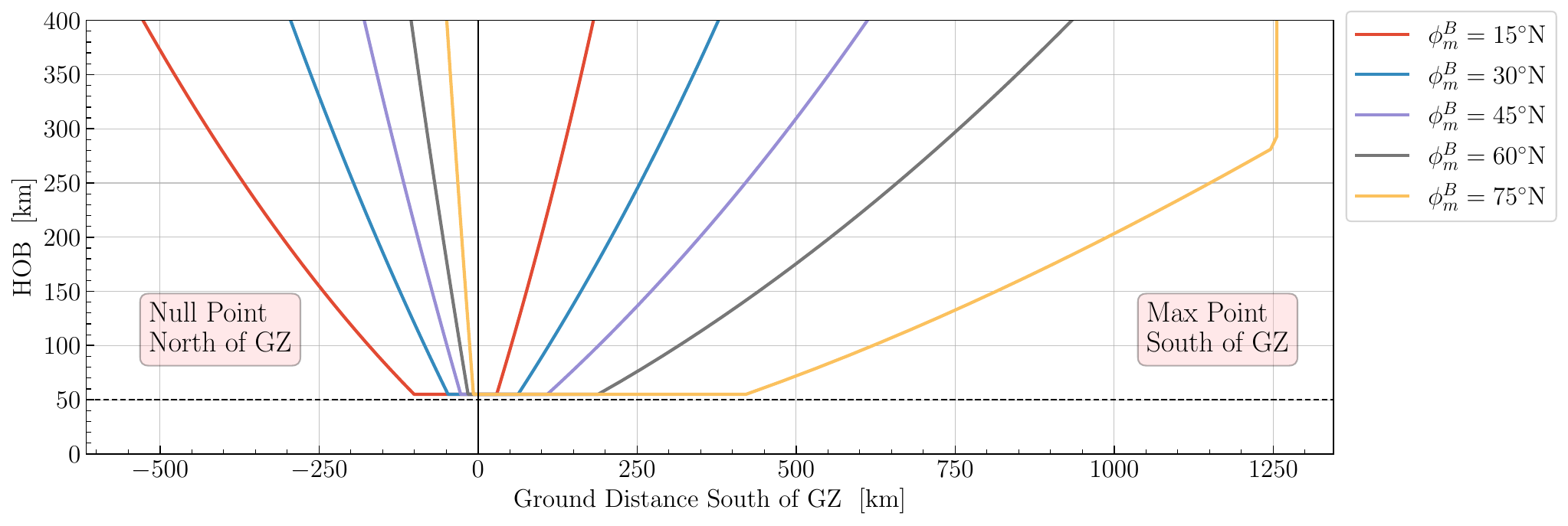}
    \caption{The ground distance, computed according to the haversine formula, from GZ to the geomagnetic null point (left side) and geomagnetic max point (right side) for bursts at different heights and  latitudes. The horizontal dashed line corresponds to the upper boundary of the absorption layer, and the nearly vertical segment for the $75^{\circ}$N-latitude burst corresponds to the fact that the max point occurs on the boundary of the line of sight cone. Note that the geomagnetic null point coincides with the EMP null point for which $E=0$, whereas the max EMP point will occur closer to GZ than the geomagnetic max point.}
	\label{fig:fig_2_11_savage_et_al}
\end{figure*}

Given that the dominant geometrical effect responsible for producing the characteristic angular pattern of the EMP surface intensity is the variation of $\theta$, following \cite{savage2010early} in Figure~\ref{fig:fig_2_11_savage_et_al} we plot the relationship between burst height and the ground distance to the geomagnetic max/null points. The geomagnetic null point is defined as the surface point for which $\theta = 0$, and the geomagnetic max point is defined as $\text{argmax} \sin\theta$. According to the above discussion, the geomagnetic null point coincides with the EMP null point for which $E = 0$, whereas the geomagnetic max point merely provides a rough estimate of the max $E$-field point. 

\begin{figure*}[!htbp]
\centering
\begin{minipage}{.48\linewidth}
    \includegraphics[width=\linewidth]{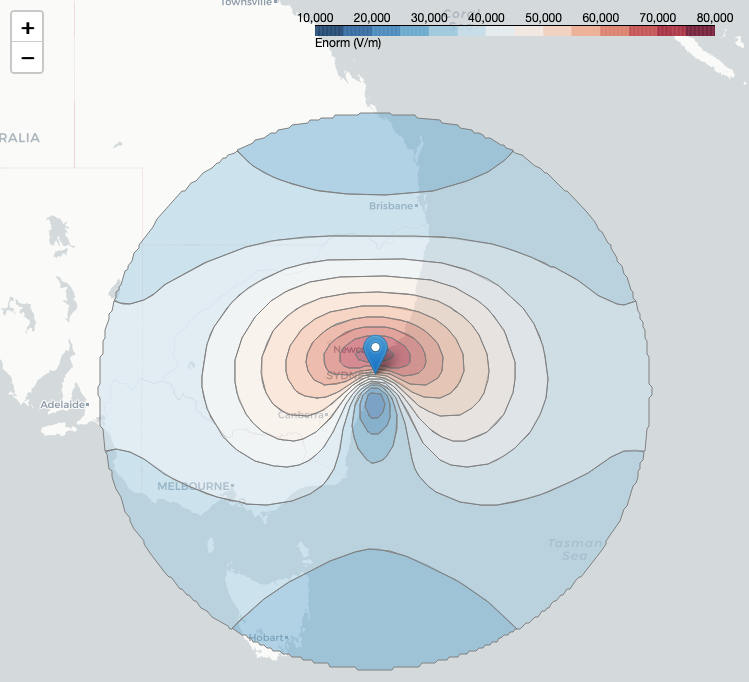}
    \caption{
Contour plot of the maximum (over time) EMP intensity for a 5 kiloton nuclear detonation 100 km directly overhead Sydney, Australia, using the dipole approximation to the geomagnetic field.
}
    \label{fig:Sydney_smile}
\end{minipage}
\hfill
\begin{minipage}{.48\linewidth}
    \includegraphics[width=\linewidth]{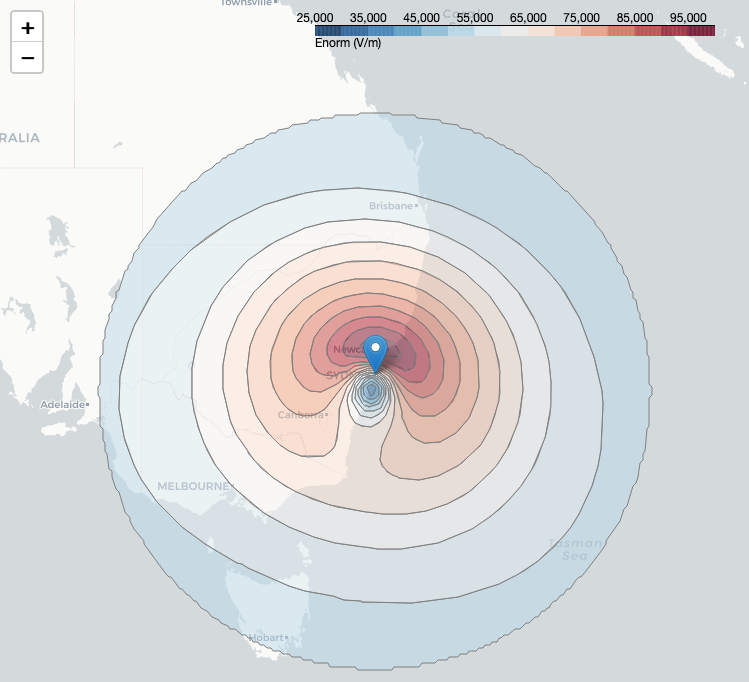}
    \caption{
Contour plot of the maximum (over time) EMP intensity for a 5 kiloton nuclear detonation 100 km directly overhead Sydney, Australia, using the IGRF approximation to the geomagnetic field.
    }
    \label{fig:Sydney_smile_igrf}
\end{minipage}
\end{figure*}

The distinctive ``smile'' pattern is hemisphere-dependent: a burst in the Southern hemisphere would lead to a ``frown'' pattern. This is because the inclination angle of the geomagnetic field is positive in the Northern hemisphere and negative in the Southern hemisphere. For example, Figure~\ref{fig:Sydney_smile} depicts the result of a burst directly overhead Sydney, Australia, assuming (as before) a dipole geomagnetic field. Of course, the actual geomagnetic field is not a perfect dipole, and the EMP intensity contours will depend on how the geomagnetic field is being modeled. Figure~\ref{fig:Sydney_smile_igrf} shows the result for a simulation that used the more accurate IGRF model of the geomagnetic field. 

\section{Discussion \label{sec:discussion}}
In this work we have reviewed the Karzas-Latter model of a high-altitude EMP, as well as a further approximation to the model introduced by Seiler. The model is a rather crude approximation to the true phenomena, but it is nevertheless quite useful because it converts an inherently $(3+1)$-dimensional problem into a $(1+0)$-dimensional problem (one spatial coordinate only). This dramatic simplification helps to facilitate an intuitive understanding of the physics, and it also greatly aids the implementation of a numerical code of the model. In conjunction with this paper, we have made publicly available a numerical code implemented in Python.

A key motivation for this work is the value to the public and policymakers in having an open and transparent EMP modeling code. For example, the code could be used for scenario planning and wargaming in future conflicts where EMP might play a role. In particular, the ability to quickly generate diagrams such as Figure~\ref{fig:Topeka_smile} for a range of weapon and geometrical parameters should be quite useful for these exercises. The code can also be used to inform the public about the qualitative effects of an EMP attack. Lastly, we also hope that this code might be a useful resource for scientists studying EMP.
We must stress the fact that the physics modeled here is just an approximation, and a rather crude one at that. The predictions are likely only correct to within an order of magnitude, and the code must not be used in any context where accuracy is of critical importance.

\subsection*{Acknowledgements}
We would like to thank Edward Geist, Don Snyder, and Andrew Lohn for useful discussions on EMP phenomenology and Alvin Moon for his careful review of the code.

The research reported here was completed in February 2023 and underwent security review with the Defense Office of Prepublication and Security Review before public release.
This research was conducted within the Acquisition and Technology Policy (ATP) Program of the RAND National Security Research Division (NSRD), which operates the RAND National Defense Research Institute (NDRI), a federally funded research and development center (FFRDC) sponsored by the Office of the Secretary of Defense, the Joint Staff, the Unified Combatant Commands, the Navy, the Marine Corps, the defense agencies, and the defense intelligence enterprise. This research was made possible by NDRI exploratory research funding that was provided through the FFRDC contract and approved by NDRI’s primary sponsor.
For more information on the RAND ATP Program, see \href{www.rand.org/nsrd/atp}{www.rand.org/nsrd/atp}.

\appendix

\section{Seiler's Approximation \label{app:seiler}}
In this Appendix, we review Seiler's approximation to the Karzas-Latter model. The numerical integrals in the source terms, Eq.~\ref{eq:current} and Eq.~\ref{eq:conductivity}, are undesirable from an implementation perspective. This is particularly true for the expression for the current, which involves two nested integrals. Seiler's approximation allows these currents to be computed analytically.

The approximation relies on the difference of exponential form for the pulse function, Eq.~\ref{eq:pulse}. First, this choice is convenient because it allows for the rate function can be computed in closed form before any approximations are made:
\begin{equation}
	g(\bm{r}) = \frac{Y_{\gamma}}{4\pi r^2 \lambda(\bm{r}) K} \exp\left( - \frac{S e^{-H/S}}{\lambda_0 \cos(A)}  \left( e^{r \cos(A)/S} - 1 \right)  \right) \,.
\end{equation}    
Second, the source term integrals may be computed analytically if an expansion around $\omega = 0$ is carried out. To leading order, the current is:
\begin{widetext}
\begin{equation}
    \sigma(\bm{r}, \tau) = 
    \begin{cases}
    0 & \tau \le 0 \,, \\
    \sigma_0(\bm{r}, \tau) \left[ \left( a \, \tau - 1 + e^{-a \, \tau} \right) \frac{b}{a} - (a \leftrightarrow b) \right] & 0 < \tau \le T \,, \\
    \sigma_0(\bm{r}, \tau) \left[ \frac{b}{a} \left( a \, T + e^{-a \tau} - a^{a(T-\tau)} \right) - (a \leftrightarrow b) \right] & \tau > T \,, 
    \end{cases}
\end{equation}

where 
\begin{equation}
    \sigma_0(\bm{r}, \tau) = \frac{e^2 q}{m} \frac{g(\bm{r})}{\nu_c(\tau)} \frac{1}{(b-a) T} \,,
\end{equation}
and where $T = (1 - \beta) \min(1\mu s , R(\bm{r})/V_0) $ is the relativistically scaled lifetime of the Compton electrons. Here we follow Seiler impose an upper limit of 1 $\mu s$ on the unscaled lifetime. The latitudinal Compton current is:
\begin{equation}
	\label{eq:currenttheta}
    j_{\theta}^C(\bm{r},\tau)
    =
    \begin{cases}
    0 & \tau \le 0 \,, \\
    j_{\theta,0}^C(\bm{r}) \left[ \left( (a \, \tau)^2 - 2 a \, \tau + 2 - 2 e^{-a \, \tau} \right) \frac{b}{a^2} - (a \leftrightarrow b) \right] & 0 < \tau \le T \,,  \\
    j_{\theta,0}^C(\bm{r}) \left[ e^{-a \, \tau} \left( e^{a \, T} \left( (a \, T)^2 - 2 a \, T + 2 \right) - 2 \right) \frac{b}{a^2} - (a \leftrightarrow b) \right] & \tau > T \,, 
    \end{cases}
\end{equation}
where
\begin{equation}
	\label{eq:current_theta_prefactor}
    j_{\theta,0}^C(\bm{r}) = e g(\bm{r}) \sin(2\theta) \frac{\omega^2}{4} \frac{V_0}{b-a} \frac{1}{(1-\beta)^3} \,,
\end{equation}
and the azimuthal Compton current is:
\begin{equation}
	\label{eq:currentphi}
    j_{\phi}^C(\bm{r},\tau)
    =
    \begin{cases}
    0 & \tau \le 0 \,, \\
    j_{\phi,0}^C(\bm{r}) \left[ \left( a \, \tau - 1 + e^{-a \, \tau} \right) \frac{b}{a} - (a \leftrightarrow b) \right] & 0 < \tau \le T \,, \\
    j_{\phi,0}^C(\bm{r}) \left[ e^{-a \, \tau} \left( e^{a \, T} (a \, T - 1) + 1 \right) \frac{b}{a} - (a \leftrightarrow b)  \right] & \tau> T \,, 
    \end{cases}
\end{equation}
where
\begin{equation}
	\label{eq:current_phi_prefactor}
    j_{\phi,0}^C(\bm{r}) = - e g(\bm{r}) \sin(\theta) \omega \frac{V_0}{b-a} \frac{1}{(1-\beta)^2} \,.
\end{equation}
\end{widetext}

\begin{figure*}[!htbp]
    \centering
    \includegraphics[width=0.95\textwidth]{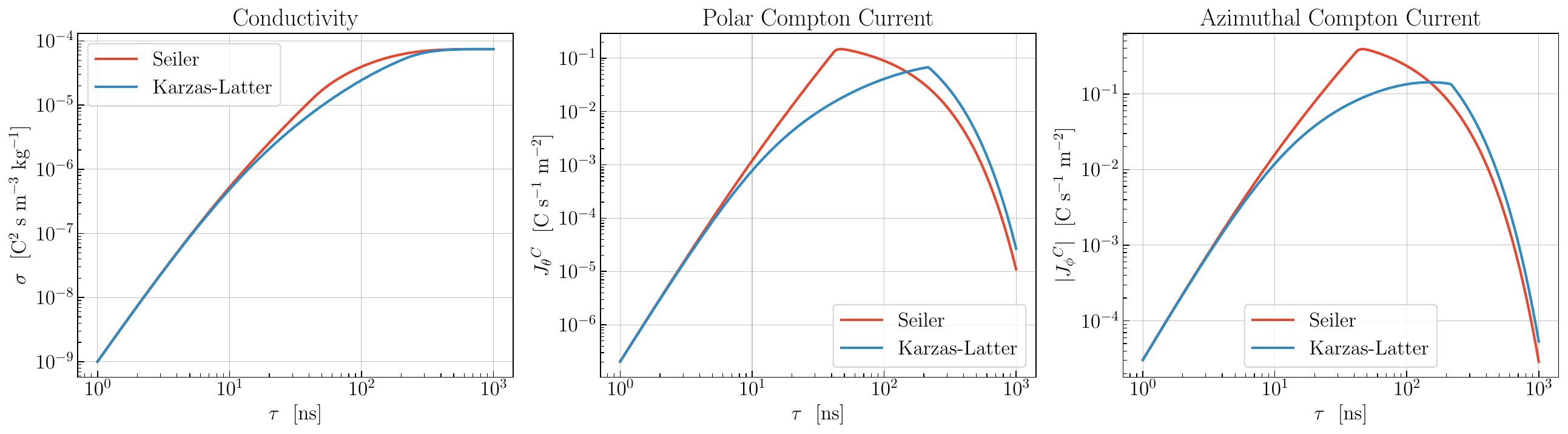}
    \caption{The conductivity (\textit{Left}), polar Compton current (\textit{Center}), and azimuthal Compton current (\textit{Right}), evaluated at the midway point between the upper and lower absorption layers, computed using both the original Karzas-Latter expressions and the Seiler approximations. For the conductivity, an electron collision frequency of $4 \times 10^3$ ns$^{-1}$ has been used. The azimuthal current is negative; here the absolute value is plotted.}
    \label{fig:Seiler_vs_KL}
\end{figure*}

This approximation is valid when the dimensionless product $\omega \tau$ is small, corresponding to early times or a weak geomagnetic field. Note that the order of approximation is different for each of the three expressions above: the conductivity is non-zero to zeroth order, the azimuthal current is non-zero to linear order, and the latitudinal current is non-zero to quadratic order. 
In Figure~\ref{fig:Seiler_vs_KL} we plot the three source terms computed using both the original Karzas-Latter equations and Seiler's approximation. The sources are plotted as a function of time for an evaluation point midway in the absorption region, for a line of sight vector with $A=0$ and $\theta = 45^{\circ}$. Each source term approaches zero as $\tau \rightarrow 0$. At late times, the conductivity approaches a constant and the currents approach zero. The quality of the Seiler approximation is seen to be poor for $\tau$ ranging from a few tens of nanoseconds to a few hundred. The fact that the Seiler approximations exhibit the correct late-time behavior implies that the absolute error will be zero as $\tau \rightarrow \infty$. 

\section{Problem Geometry and the Dipole Approximation to the Geomagnetic Field \label{app:geomagnetic}}
Included in the inputs to the Karzas-Latter model are $A$, the angle between the line of sight vector and the vertical, $\theta$, the angle between the line of sight vector and the geomagnetic field (assumed constant along the line of sight), and $B_E$, the norm of the geomagnetic field. In this Appendix, we provide details on how to compute these quantities as a function of burst point $\mathcal{B}$ and target point $\mathcal{T}$ for the case where the geomagnetic field has been approximated as an ideal dipole and the Earth has been approximated as a perfect sphere. We note that the code also supports more accurate treatment of the geomagnetic field through the most recent International Geomagnetic Reference Field (IGRF) models, which are based on a spherical harmonic expansion up to and including degree 10 or 13, depending on the year \cite{alken2021international}. IGRF support has been made possible through the use of the ppigrf package \cite{ppigrf}. 

\subsection{Coordinate Systems}
First, introduce a geographic coordinate system centered around an assumed spherical Earth, $(r_g, \phi_g, \lambda_g)$, with $r_g \ge 0$, $\phi_g \in [-\pi/2, \pi/2]$, $\lambda_g \in [-\pi, \pi]$. These can be mapped to Cartesian coordinates $(x_g, y_g, z_g)$ via \cite{hapgood1992space}:
\begin{subequations}
	 \begin{equation} x_g = r_g \cos\phi_g \cos\lambda_g \,, \end{equation}
	 \begin{equation}y_g = r_g \cos\phi_g \sin\lambda_g \,, \end{equation}
	 \begin{equation} z_g = r_g \sin\phi_g \,. \end{equation}
\end{subequations}
The inverse map is:
\begin{subequations}
	 \begin{equation} \phi_g = \tan^{-1}\left( \frac{z_g}{\sqrt{x_g^2 + y_g^2}} \right) \,, \end{equation}
	 \begin{equation}
		 \lambda_g = 
		\begin{cases}
		\cos^{-1}\left( \frac{x_g}{\sqrt{x_g^2 + y_g^2}} \right) \,,& y_g > 0 \,, \\
		- \cos^{-1}\left( \frac{x_g}{\sqrt{x_g^2 + y_g^2}} \right) \,, & y_g \le 0 \,.
		\end{cases}
	\end{equation}
\end{subequations}
The unit vectors associated with this coordinate system are:
\begin{subequations}
	\begin{equation}
		\hat{ \bm{r} }_g = \cos\phi_g \cos\lambda_g \, \hat{ \bm{x} }_g + \cos\phi_g \sin\lambda_g \, \hat{ \bm{y} }_g + \sin\phi_g \, \hat{ \bm{z} }_g
	\end{equation}
	\begin{equation}
		\hat{ \bm{\phi} }_g = -\sin\phi_g \cos\lambda_g \, \hat{ \bm{x} }_g - \sin\phi_g \sin\lambda_g \, \hat{ \bm{y} }_g + \cos\phi_g \, \hat{ \bm{z} }_g
	\end{equation}
	\begin{equation}
		\hat{ \bm{\lambda} }_g = -\sin\lambda_g \, \hat{ \bm{x} }_g + \cos\lambda_g \, \hat{ \bm{y} }_g	
	\end{equation}
\end{subequations}
The $g$ subscript refers to the fact that the $z$-axis is aligned with the Earth's rotational axis. 

For the purposes of describing the geomagnetic field, it will be convenient to introduce a second set magnetic of coordinates $(r_m, \phi_m, \lambda_m)$ with the $z$-axis aligned with the dipole moment. (Note that these ``magnetic coordinates'' are similar to, but not the same as the ``geomagnetic coordinate system'' \cite{hapgood1992space}.). The associated Cartesian coordinates are
\begin{subequations}
	\begin{equation} x_m = r_m \cos\phi_m \cos\lambda_m \,, \end{equation} 
	\begin{equation} y_m = r_m \cos\phi_m \sin\lambda_m \,, \end{equation}
	\begin{equation} z_m= r_m \sin\phi_m \,. \end{equation}
\end{subequations}

The geographic coordinates will be used for the burst and target points, whereas the magnetic coordinates will be used to evaluate the geomagnetic field. The model parameter $\theta$ is the angle between the line of sight vector and the geomagnetic vector, and therefore it is necessary to transform between the two sets of coordinates. The two Cartesian coordinates differ by the choice of $z$-axis - and hence, by a rotation:
\begin{equation}
	\bm{x}_g = \bm{R}_{\bm{v}}(\varphi) \bm{x}_m \,.
\end{equation}
Here, $\bm{R}_{\bm{v}}(\varphi)$ is the rotation matrix corresponding to a rotation of $\varphi$ radians around the axis specified by the vector $\bm{v}$, and is given by the Rodrigues formula:
\begin{widetext}
\begin{equation}
	\label{eq:rodrigues}
	\bm{R}_{\bm{v}}(\varphi) = 
	\begin{pmatrix}
	\cos\varphi + v_x^2 (1-\cos\varphi) & v_x v_y (1-\cos\varphi) - v_z \sin\varphi & v_x v_z (1 - \cos\varphi) + v_y \sin\varphi \\
	v_x v_y (1-\cos\varphi) + v_z \sin\varphi & \cos\varphi + v_y^2 (1-\cos\varphi) & v_y v_z (1-\cos\varphi) - v_x \sin\varphi \\
	v_x v_z (1-\cos\varphi) - v_y \sin\varphi & v_y v_z (1-\cos\varphi) + v_x \sin\varphi & \cos\varphi + v_z^2 (1-\cos\varphi) \,.
	\end{pmatrix}
\end{equation}
\end{widetext}
The conversion from one set of angular coordinates to another may be accomplished by first mapping to the corresponding Cartesian coordinates, applying the rotation matrix (or its inverse), and then mapping back to angular coordinates. It only remains to specify the angle $\varphi$ and axis $\bm{v}$ of rotation. 

The rotation matrix should map the magnetic North Pole (mNP), located at $(x_m^{\text{mNP}}, y_m^{\text{mNP}}, z_m^{\text{mNP}}) = (0, 0, 1)$, to its location in geographic coordinates, which is approximately 86.294°N, 151.948°E as of 2022 \cite{NOAA}, corresponding to $(\phi_g^{\text{mNP}}, \lambda_g^{\text{mNP}}) = (1.506, 2.652)$ rad.
This can be accomplished with rotation parameters $\varphi = \pi/2 -\phi_g^{\text{mNP}}$ and $\bm{v} = (-\sin\lambda_g^{\text{mNP}}, \cos\lambda_g^{\text{mNP}}, 0)$; substiting these values in Eq.~\ref{eq:rodrigues} yields:
\begin{equation}
	\begin{pmatrix}
		x_g^{\text{mNP}} \\
		y_g^{\text{mNP}} \\
		z_g^{\text{mNP}} 
	\end{pmatrix}
	= 
	\begin{pmatrix}
		v_y \sin\varphi \\
		-v_x \sin\varphi \\
		\cos\varphi 
	\end{pmatrix}
	=	
	\begin{pmatrix}
		\cos\phi_g^{\text{mNP}} \cos\lambda_g^{\text{mNP}} \\
		\cos\phi_g^{\text{mNP}} \sin\lambda_g^{\text{mNP}} \\
		\sin\phi_g^{\text{mNP}}
	\end{pmatrix}	
\end{equation}
as desired. 

\subsection{Spatial Variation of Model Parameters}
Equipped with these coordinate systems and the transformation maps, we now describe how to compute $B_E$, $A$, and $\theta$. Given the many approximations made up to this point, it seems reasonable to model the geomagnetic field as a simple magnetic dipole centered around a spherical Earth:
\begin{subequations}
	\label{eq:dipole}
    \begin{equation} (B_E)_{r_m} = - 2 B_0 \left( \frac{R_E}{r_m} \right)^3 \sin\phi_m \,, \end{equation}
     \begin{equation} (B_E)_{\phi_m} = B_0 \left( \frac{R_E}{r_m} \right)^3 \cos\phi_m \,, \quad \end{equation}
     \begin{equation} (B_E)_{\lambda_m} = 0 \,, \end{equation}
\end{subequations}
(The overall minus sign is due to the fact that the Earth's magnetic ``North Pole'' is actually the South Pole of the dipole.) The total magnetic field strength is $B_E = B_0 \left( \frac{R_E}{r_m} \right)^3 \sqrt{1 + 3 \sin^2\phi_m}$, which changes by a factor of 2 over the surface of the Earth, with maxima at the (magnetic) poles and minima at the (magnetic) equator. Although the geomagnetic field varies continuously with position, when performing the line of sight integration to solve the sourced Maxwell equations (Eq.~\ref{eq:maxwell}) we assume that the field is constant, with the specific value corresponding to the midway point between the upper and lower absorption layers along the line of sight.

Next, the angle $A$ may be computed for each pair of burst and target points, $\mathcal{B}$, $\mathcal{T}$, with coordinates:
\begin{equation}
    \mathcal{T} = (R_E, \lambda^{\mathcal{T}}, \phi^{\mathcal{T}}) \,, \quad \mathcal{B} = (R_E+H, \lambda^{\mathcal{B}}, \phi^{\mathcal{B}}) \,.
\end{equation}
The vector pointing from the burst to the target is then $\boldsymbol{x}_{\mathcal{B}\mathcal{T}} = \boldsymbol{x}_{\mathcal{T}} - \boldsymbol{x}_{\mathcal{B}}$ and the vector pointing from the burst to the origin is $-\boldsymbol{x}_{\mathcal{B}}$. Using these, the angle $A$ satisfies
\begin{equation}
    \cos A = - \frac{\boldsymbol{x}_{\mathcal{B}\mathcal{T}} \cdot \boldsymbol{x}_{\mathcal{B}}}{|\boldsymbol{x}_{\mathcal{B}\mathcal{T}}| |\boldsymbol{x}_{\mathcal{B}}|} \,.
\end{equation}

Similarly, the angle $\theta$ may be obtained using
\begin{equation}
	\cos\theta = \frac{\boldsymbol{x}_{\mathcal{B}\mathcal{T}} \cdot \boldsymbol{B}_E} {|\boldsymbol{x}_{\mathcal{B}\mathcal{T}}| |\boldsymbol{B}_E|} \,.
\end{equation}
Note that in evaluating this expression care should be taken to first transform the magnetic coordinates used to express $\bm{B}_E$ into geographic coordinates.

\bibliography{refs}
\bibliographystyle{JHEP}

\end{document}